\documentclass{elsart}
\usepackage{epsfig}
\usepackage{amssymb}

\begin{document}
\parindent=1em

\begin{frontmatter}

\title{Proton-nucleus elastic scattering and the equation of state
of nuclear matter} 

\author[a]{Kei Iida},
\ead{keiiida@riken.go.jp}
\author[a,b]{Kazuhiro Oyamatsu},
\author[a,c]{Badawy Abu-Ibrahim}

\address[a]{The Institute of Physical and Chemical Research (RIKEN),
Hirosawa, Wako, Saitama 351-0198, Japan}

\address[b]{Department of Media Theories and Production, Aichi Shukutoku
University, Nagakute, Nagakute-cho, Aichi-gun, Aichi 480-1197, Japan}

\address[c]{Department of Physics, Cairo University, Giza 12613, Egypt}

\begin{abstract}

     We calculate differential cross sections for proton-nucleus elastic
scattering by using a Glauber theory in the optical limit approximation and
nucleon distributions that can be obtained in the framework of macroscopic 
nuclear models in a way dependent on the equation of state of uniform nuclear 
matter near the saturation density.  We find that the peak angle calculated 
for unstable neutron-rich nuclei in the small momentum transfer regime 
increases as the parameter $L$ characterizing the density dependence of the 
symmetry energy decreases.  This is a feature associated with the $L$ 
dependence of the predicted matter radii.

\end{abstract}
\end{frontmatter}

\newpage

     The equation of state (EOS) of nuclear matter is essential to 
understanding of the saturation property of atomic nuclei \cite{I}, the 
structure of neutron stars \cite{HP}, the mechanism of stellar collapse
\cite{L}, and the dynamics of relativistic heavy-ion collisions \cite{DLL}.  
Generally, the energy per nucleon of nuclear matter can be expanded around the
saturation point of symmetric nuclear matter as
\begin{equation}
    w=w_0+\frac{K_0}{18n_0^2}(n-n_0)^2+ \left[S_0+\frac{L}{3n_0}(n-n_0)
      \right]\alpha^2.
\label{eos0}
\end{equation}
Here $w_0$, $n_0$ and $K_0$ are the saturation energy, the saturation density 
and the incompressibility of symmetric nuclear matter, $n$ is the nucleon
density, and $\alpha=1-2x$ is the neutron excess.  The parameters $L$ and 
$S_0$ characterize the density dependent symmetry energy coefficient $S(n)$: 
$S_0$ is the symmetry energy coefficient at $n=n_0$, and 
$L=3n_0(dS/dn)_{n=n_0}$ is the symmetry energy 
density derivative coefficient (hereafter referred to as the ``density 
symmetry coefficient'').  As shown in Ref.\ \cite{I} by describing macroscopic
nuclear properties in a manner that is dependent on these EOS parameters, 
empirical data for masses and radii of stable nuclei can provide a strong 
constraint on the parameters $w_0$, $n_0$ and $S_0$, while leaving $K_0$ and 
$L$ uncertain.  We remark that the isoscalar giant monopole resonance in 
nuclei (e.g., Ref.\ \cite{YCL}) and caloric curves in nuclear collisions 
(e.g., Ref.\ \cite{Nat}) can constrain $K_0$ only in a way that is dependent 
on models for the effective nucleon-nucleon interaction.

     The incompressibility $K_0$ and the density symmetry coefficient $L$ 
control in which direction the saturation point moves on the density versus 
energy plane, as the neutron excess increases from zero.  This feature can be 
found from the fact that up to second order in $\alpha$, the saturation energy
$w_s$ and density $n_s$ are given by 
\begin{equation}
  w_s=w_0+S_0 \alpha^2
\label{ws}
\end{equation}
and
\begin{equation}
  n_s=n_0-\frac{3 n_0 L}{K_0}\alpha^2.
\label{ns}
\end{equation}
The influence of $K_0$ and $L$ on neutron star structure can be significant
\cite{Prakash} despite the fact that these parameters characterize
the EOS near normal nuclear density and proton fraction, which are fairly 
small and large, respectively, as compared with the typical densities and
proton fractions in the central region of the star.

     In our previous investigation \cite{I} we pointed out the possibility that
future systematic measurements of the matter radii of unstable neutron-rich 
nuclei help derive the value of $L$.  As an example, the matter radii of 
unstable neutron-rich nuclei such as Ni and even heavier element isotopes are 
expected to be deduced from future measurements of proton-nucleus elastic 
differential cross sections that will be performed by using a beam of such 
nuclei incident on a proton target and detecting scattered protons.  This 
possibility is supported by our finding based on macroscopic nuclear models, 
which reasonably reproduce empirical data for masses and radii of stable 
nuclei and allow for uncertainties in the values of $K_0$ and $L$, that the 
matter radii calculated for unstable nuclei depend appreciably on $L$, while 
being almost independent of $K_0$.  However, extraction of the matter radii 
from experimental data for proton-nucleus elastic differential cross sections 
and for interaction cross sections is not straightforward in the sense that it
requires the approximate scattering theory \cite{Batty,Ray}.  It is thus 
instructive to examine how the cross sections themselves are related to the 
parameter $L$ in a proper theoretical framework.

     In this paper we focus on proton-nucleus elastic scattering and, in the 
optical limit approximation of the Glauber multiple scattering model \cite{G},
obtain its angular distribution from the nucleon distributions in a nucleus 
calculated in Ref.\ \cite{I} in a way dependent on $L$ and $K_0$.  For 
sufficiently high proton incident energies and small momentum tranfers to 
validate the Glauber model, we find that the angle of the scattering peak
decreases with $L$ more remarkably for larger neutron excess, while the peak 
height increases with $K_0$ almost independently of neutron excess.  We 
suggest the possibility that comparison of the calculations with experimental 
data for the peak angle may be useful for determination of $L$.

      We begin with macroscopic nuclear models used in this and a previous
work \cite{I}.  We describe a spherical nucleus of proton number $Z$ and mass 
number $A$ within the framework of a simplified version of the extended 
Thomas-Fermi theory \cite{O}.  We first write the total energy of a nucleus as
a function of the neutron and proton density distributions $n_n({\bf r})$ and 
$n_p({\bf r})$ in the form
\begin{equation}
 E=E_b+E_g+E_C+(A-Z)m_n c^2+Zm_p c^2,
\label{e}
\end{equation}
where 
\begin{equation}
  E_b=\int d^3 r n({\bf r})w\left(n_n({\bf r}),n_p({\bf r})\right)
\label{eb}
\end{equation}
is the bulk energy with the energy per nucleon $w(n_n,n_p)$ of uniform nuclear
matter,
\begin{equation}
  E_g=F_0 \int d^3 r |\nabla n({\bf r})|^2
\label{eg}
\end{equation}
is the gradient energy with adjustable constant $F_0$,
\begin{equation}
  E_C=\frac{e^2}{2}\int d^3 r \int  d^3 r' 
      \frac{n_p({\bf r})n_p({\bf r'})}{|{\bf r}-{\bf r'}|}
\label{ec}
\end{equation}
is the Coulomb energy, and $m_n$ and $m_p$ are the neutron and proton rest
masses.  We express $w$ as \cite{O}
\begin{equation}
  w=\frac{3 \hbar^2 (3\pi^2)^{2/3}}{10m_n n}(n_n^{5/3}+n_p^{5/3})
      +(1-\alpha^2)v_s(n)/n+\alpha^2 v_n(n)/n,
\label{eos1}
\end{equation}
where 
\begin{equation}
  v_s=a_1 n^2 +\frac{a_2 n^3}{1+a_3 n}
\label{vs}
\end{equation}
and
\begin{equation}
  v_n=b_1 n^2 +\frac{b_2 n^3}{1+b_3 n}
\label{vn}
\end{equation}
are the potential energy densities for symmetric nuclear matter and pure 
neutron matter.  This expression for $w$ is one of the simplest 
parametrization that reduces to Eq.\ (\ref{eos0}) in the simultaneous limit of
$n\to n_0$ and $\alpha\to0$.  We then set the nucleon distributions $n_i(r)$ 
$(i=n,p)$ as
\begin{equation}
  n_i(r)=\left\{ \begin{array}{lll}
  n_i^{\rm in}\left[1-\left(\displaystyle{\frac{r}{R_i}}\right)^{t_i}\right]^3,
         & \mbox{$r<R_i,$} \\
             \\
         0,
         & \mbox{$r\geq R_i,$}
 \end{array} \right.
\label{ni}
\end{equation}
and in the spirit of the Thomas-Fermi approximation minimize the total 
energy $E$ with respect to $R_i$, $t_i$ and $n_i^{\rm in}$ with the mass 
number $A$, the EOS parameters $(n_0,w_0,S_0,K_0,L)$ and the gradient 
coefficient $F_0$ fixed.  By calculating the charge number, mass excess and 
root-mean-square charge radius from the minimizing values of $R_i$, $t_i$ and 
$n_i^{\rm in}$ and fitting the results to the empirical values for stable 
nuclei $(25\leq A \leq 245)$ on the smoothed beta stability line, we finally 
obtain $n_0$, $w_0$, $S_0$ and $F_0$ for various sets of $L$ and $K_0$.  We 
remark that as a result of this fitting, the parameters $a_1$--$b_2$ become 
functions of $K_0$ and $L$, while we fix the remaining parameter $b_3$, which
controls the EOS of matter for large neutron excess and high density, at 
1.58632 fm$^3$ throughout this fitting process.

     The macroscopic nuclear models used here can describe gross nuclear 
properties such as masses and root-mean-square radii in a manner that is 
dependent on the EOS parameters, $L$ and $K_0$.  Notably, as in Ref.\ \cite{I},
these models predict that the root-mean-square matter radii depend appreciably
on $L$, while being almost independent of $K_0$.  However, there are some 
limitations in the present macroscopic approach.  First, this approach works 
well in the range of $\alpha\lesssim0.3$ and $A\gtrsim50$, where a macroscopic
 view of the system is relevant.  Second, the nuclear surface is not 
satisfactory in the present Thomas-Fermi-type theory, which tends to 
underestimate the surface diffuseness and does not allow for the tails of the 
nucleon distributions.  As we shall see, such an underestimated diffuseness 
has consequence to calculations of proton-nucleus differential cross sections.
Lastly, no shell and pairing effects are included.

    We proceed to calculate differential cross sections for proton elastic
scattering off nuclei using the Glauber theory in the optical limit 
approximation.  The elastic differential cross section at given momentum 
transfer ${\bf q}$ and incident proton energy $T_p$ can be written as (e.g., 
Ref.\ \cite{Ahmad})
\begin{equation}
\frac{d\sigma}{d\Omega}=|F({\bf q})|^2,
\end{equation}
with the elastic scattering amplitude,
\begin{equation}
|F({\bf q})|=\left|F_{\rm C}({\bf q})+\frac{ik}{2\pi}\int d{\bf b}
e^{-i{\bf q\cdot b}+2i\eta\ln(k|{\bf b}|)}
\left[1-e^{i\chi_{\rm N}({\bf b})}\right]\right|.
\label{amp}
\end{equation}
Here, ${\bf b}$ is the impact parameter, $\hbar k=\sqrt{(T_p/c+m_p c)^2-
(m_p c)^2}$ is the incident proton momentum, $\eta=Ze^2/\hbar v$ with the 
incident proton velocity $v=\hbar k c/(T_p/c+m_p c)$ is the Sommerfeld 
parameter, 
\begin{equation}
F_{\rm C}({\bf q})=-\frac{2\eta k}{{\bf q}^2}
     \exp\left[-2i\eta\ln\left(\frac{|{\bf q}|}{2k}\right)
       +2i {\rm arg}\Gamma(1+i\eta)\right]
\end{equation}
is the amplitude of the Coulomb elastic scattering, which we approximate as 
a usual Rutherford scattering off a point charge, and 
\begin{equation}
i\chi_{\rm N}({\bf b})=-\int d{\bf r} [n_p({\bf r})\Gamma_{pp}({\bf b}-{\bf s})
+n_n({\bf r})\Gamma_{pn}({\bf b}-{\bf s})]
\label{phase}
\end{equation}
is the phase shift function with the projection ${\bf s}$ of the coordinate
${\bf r}$ on a plane perpendicular to the incident proton momentum and with
the profile function $\Gamma_{pN}$ of the free proton-nucleon ($pN$) 
scattering amplitude, for which we use a simple parametrization,
\begin{equation}
\Gamma_{pN}({\bf b})=\frac{1-i\alpha_{pN}}{4\pi\beta_{pN}}\sigma_{pN}
                 \exp(-{\bf b}^2/2\beta_{pN}),
\label{profile}
\end{equation}
where $\sigma_{pN}$ is the $pN$ total cross section, $\alpha_{pN}=-{\rm Im}
\Gamma_{pN}(0)/{\rm Re}\Gamma_{pN}(0)$, and $\beta_{pN}$ is the slope 
parameter.

     In calculating the differential cross section, we use a numerical code 
based on the Monte Carlo integration for evaluations of the phase shift 
function (\ref{phase}), which can be applied to an arbitrary form of nucleon 
distributions.  The essential input in such calculations is a set of the
parameters, $\alpha_{pN}$, $\beta_{pN}$ and $\sigma_{pN}$, that characterize 
the profile function (\ref{profile}).  Here the values of $\alpha_{pN}$, 
$\beta_{pN}$ and $\sigma_{pN}$ at given incident proton energy $T_p$ are taken
from Ref.\ \cite{Ray1}.  We finally obtain the differential cross section as a
function of the C.M. scattering angle, $\theta_{\rm c.m.}$, from the nucleon 
distributions (\ref{ni}) in a nucleus determined as functions of the 
parameters $L$ and $K_0$ in the macroscopic nuclear models.

\begin{figure}
\begin{center}
\epsfig{file=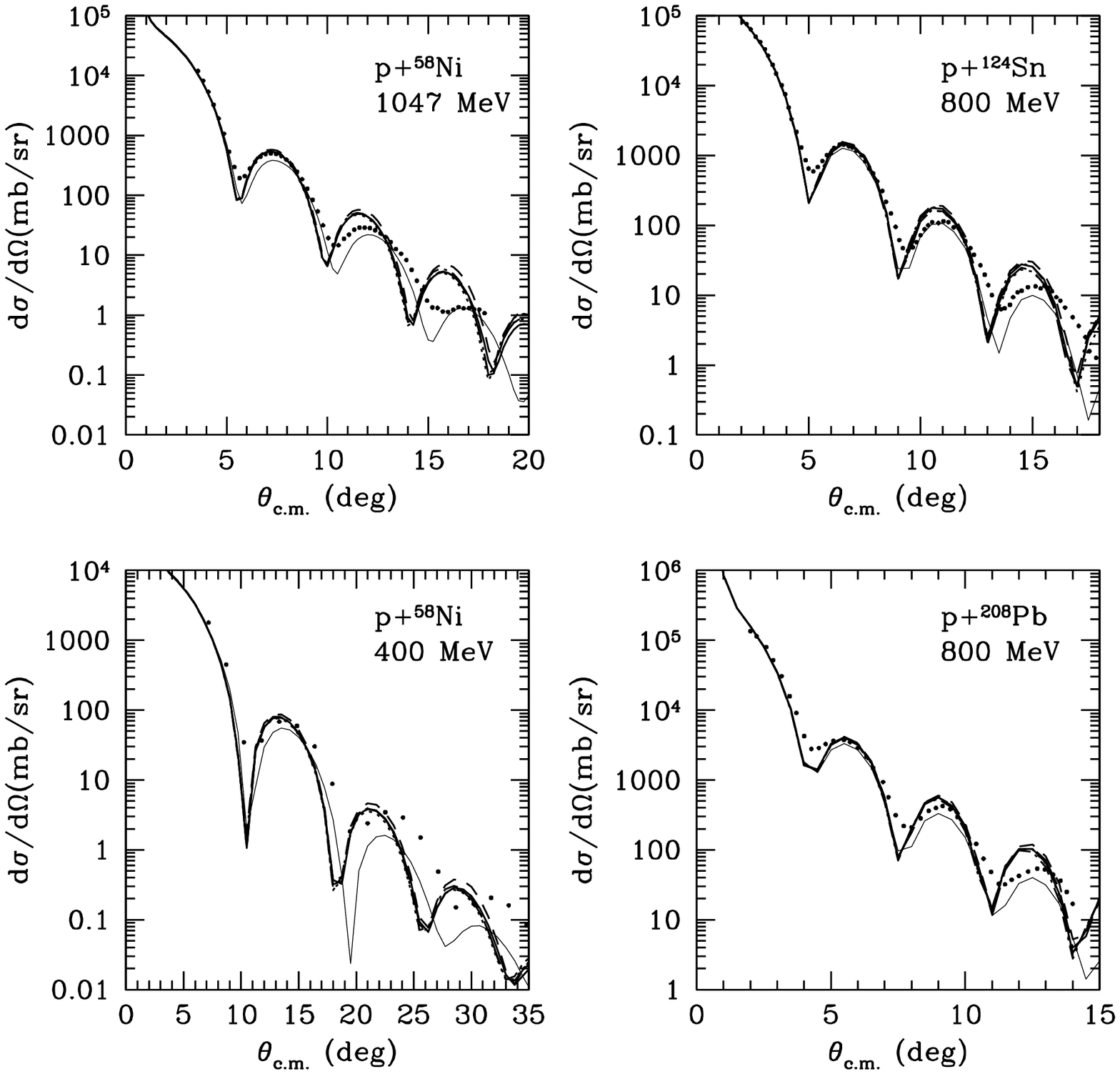,height=13cm}
\end{center}

Fig.\ 1.  Proton-nucleus elastic differential cross sections for $^{58}$Ni 
($T_p=1047,$ 400 MeV), $^{124}$Sn ($T_p=800$ MeV) and $^{208}$Pb ($T_p=800$ 
MeV).  The solid, dotted, short-dashed, long-dashed and dot-dashed lines are
the results calculated for $(L,K_0)=(50,230), (50,180), (50,360), (5,230), 
(80,230)$ in MeV.  For comparison, we also plotted the results (thin lines) 
calculated from the Fermi-type nucleon distributions derived in Ref.\ 
\cite{Ray2}.  The experimental data (dots) are taken from Refs.\ 
\cite{Ray2,Lom,Saka}.

\end{figure}

     We start with proton elastic scattering off stable nuclei.  Figure 1 
displays the elastic differential cross sections obtained for
$^{58}$Ni ($T_p=1047, 400$ MeV), $^{124}$Sn ($T_p=800$ MeV), and $^{208}$Pb
($T_p=800$ MeV).  In each panel the calculations were performed at $(L,K_0)
=$(5,230), (50,230), (80,230), (50,180), (50,360) in MeV.  The dependence of
the obtained differential cross sections on the EOS parameters is not 
appreciable.  Note that the present Glauber model is valid for sufficiently 
large $T_p$ to justify the optical limit approximation and for sufficiently 
small angles to allow us to ignore the nucleon-nucleon correlations in a 
nucleus.  By comparison with the empirical data, we may conclude that the 
Glauber model reasonably describes proton-nucleus elastic scattering for high 
incident energies ($T_p\gtrsim500$ MeV) and for small scattering angles 
($\theta_{\rm c.m.}\lesssim10$ deg).

     Generally \cite{ADL}, the peak angles are related to the nuclear radius, 
while the peak heights are related to the diffuseness of the nuclear surface.
Our nuclear model, on the other hand, predicts that the radius increases with 
$L$, while the diffuseness decreases with $K_0$.  It is thus important to 
investigate the detailed peak structure in the small scattering angle regime 
and its relation with $L$ and $K_0$ within the present theoretical framework.

     In Fig.\ 2, the angles and heights of the first scattering peak, 
calculated for 228 sets of $L$ and $K_0$ ranging $0<L<175$ MeV and 180 MeV 
$\leq K_0 \leq 360$ MeV, are plotted for proton elastic scattering off stable
nuclei, $^{116}$Sn and $^{124}$Sn, at $T_p=800$ MeV.\footnote{In this paper, 
we define the zeroth peak as that whose angle corresponds to $\theta_{\rm 
c.m.}=0$.}  We first note that the peak angle decreases with $L$.  This 
correlation is larger for larger neutron excess, since the matter radii 
increase with $L\alpha^2$ \cite{I}.  Second, we find that the peak height 
increases with $K_0$ in a way almost independent of neutron excess.  We remark
in passing that there is no appreciable correlation between the peak angle and
$K_0$ as well as between the peak height and $L$.  This absence of appreciable
correlation between the peak angle and $K_0$ is consistent with 
the fact that within the present Glauber model, the Fermi-type nucleon 
distributions derived by Ray et al.\ \cite{Ray2}, which have larger surface 
diffuseness than those used in the present calculations while having similar 
root-mean-square neutron and proton radii, provide the angle of the first 
scattering peak similar to our results (see Fig.\ 1).  We also note that $K_0$
($L$) induced uncertainties in the peak angle (height) are typically
$\pm 0.03$ deg ($\pm 3$ \%), to which numerical errors due to the Monte Carlo 
integration are confined.

\begin{figure}
\begin{center}
\epsfig{file=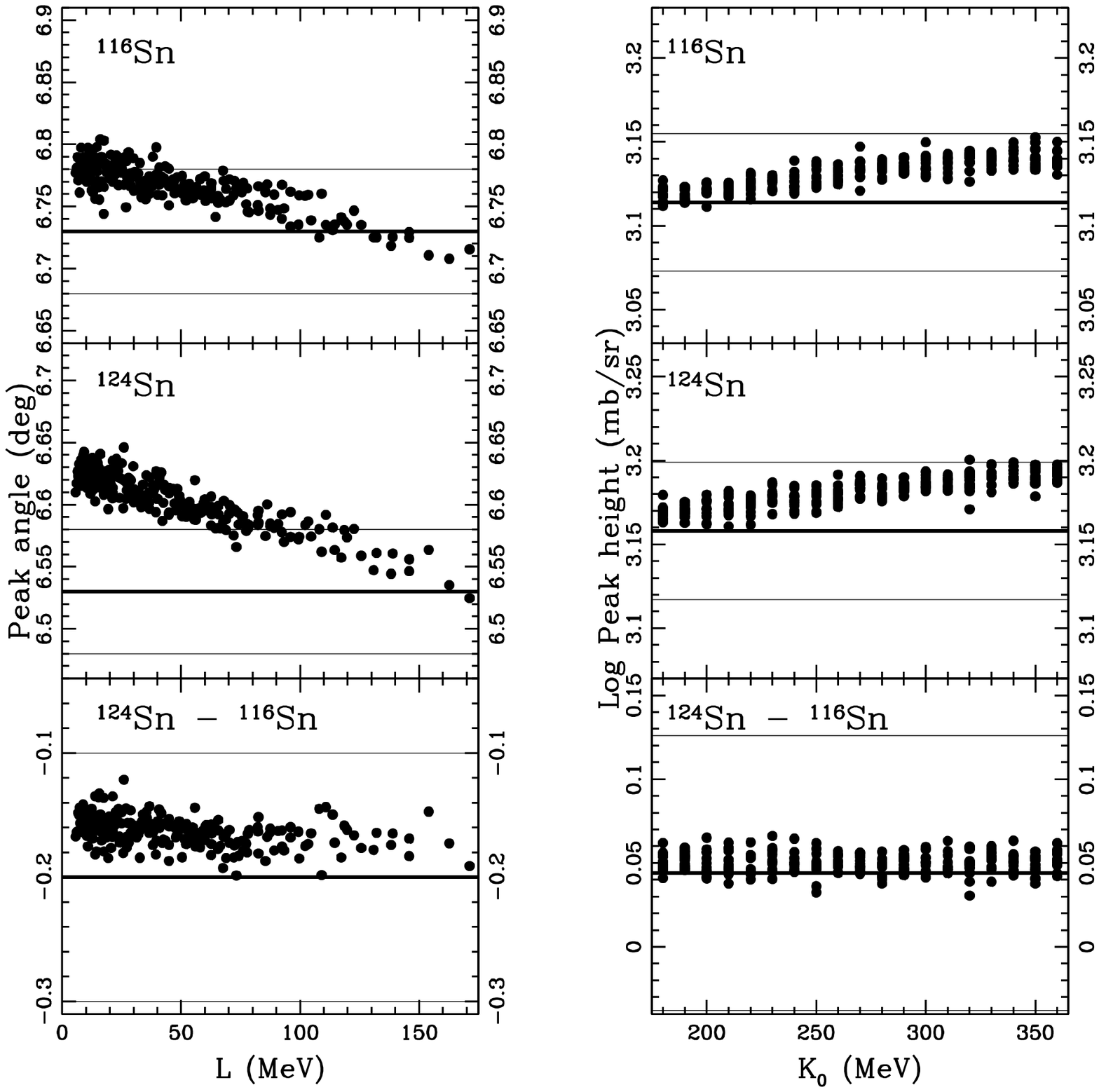,height=13cm}
\end{center}

Fig.\ 2.   The angles and heights of the scattering peak in the small angle
regime, calculated as functions of $L$ and $K_0$ for $p$-$^{116}$Sn and 
$p$-$^{124}$Sn elastic scattering at $T_p=800$ MeV.  The experimental angles 
and heights including errors (from Ref.\ \cite{Ray2}) are denoted by the
horizontal lines (thick lines: central values, thin lines: upper and lower
bounds).

\end{figure}

     The empirical data for the peak angle and height, also plotted in Fig.\ 
2, contain uncertainties in the absolute scattering angle and normalization
\cite{Ray2}, which are here taken to be $\pm0.05$ deg and $\pm10$ \%, 
respectively.  The comparison between the calculated and experimental results 
for the peak angles suggests a possible extraction of $L$.  Since the 
calculations ignore the pairing and shell effects and the tails of the nucleon
distributions, however, the comparison of the absolute values of the peak 
angles is accompanied by systematic errors.  These errors are expected to be 
reduced by considering a difference in the peak angle between the $^{116}$Sn 
and $^{124}$Sn cases.  The calculated values of this difference depend on $L$
only weakly and are within the uncertainties in the experimental value.
We remark that extraction of $K_0$ from the peak height is difficult, since 
the present calculations underestimate the surface diffuseness and the values 
of $\alpha_{pN}$, which control the peak-to-valley ratios \cite{Ray2}, have 
yet to be determined uniquely.

\begin{figure}
\begin{center}
\epsfig{file=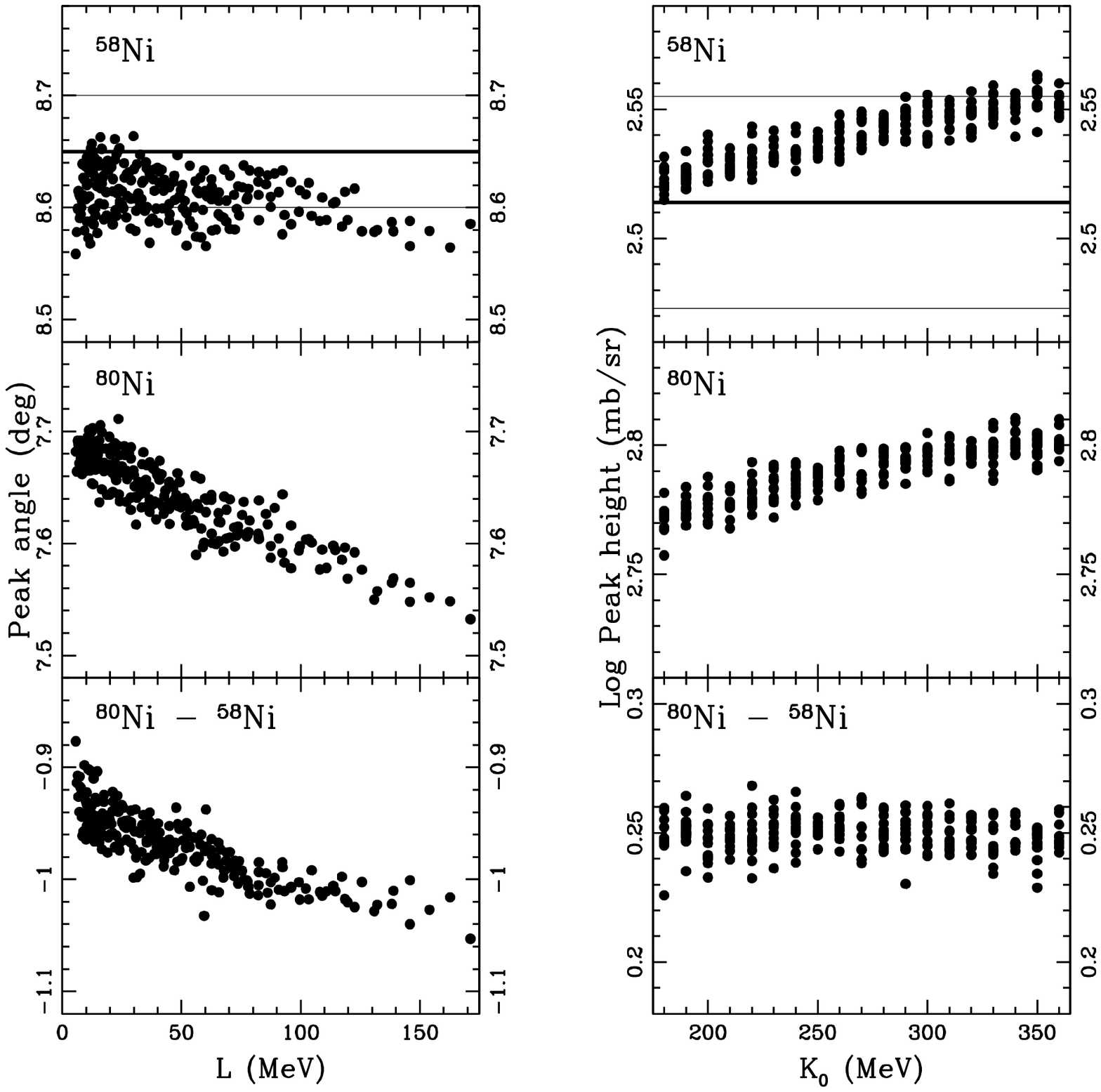,height=13cm}
\end{center}

Fig.\ 3.  Same as Fig.\ 2 for $p$-$^{58}$Ni and $p$-$^{80}$Ni elastic 
scattering at $T_p=800$ MeV.

\end{figure}

     We turn to unstable nuclei, whose beams incident on proton targets may 
provide elastic scattering data in future experiments.  In order to examine
the possibility of extracting $L$ from the peak angles for unstable nuclei
such as $^{80}$Ni ($\alpha=0.3$), we repeat the calculations for $p$-$^{58}$Ni
and $p$-$^{80}$Ni elastic scattering at $T_p=800$ MeV; the results for the
angles and heights of the first scattering peak are exhibited in 
Fig.\ 3.  We find that the $L$ dependence of the difference in the peak angle 
between the $^{58}$Ni and $^{80}$Ni cases is appreciably large.  The $L$
induced change in the angle difference, $\Delta \theta_{\rm c.m.}$, amounts to
order 0.1 deg.  This is related to the $L$ induced change in the matter radius
difference, $\Delta R_{\rm m} \sim 0.1$ fm \cite{I}, by the diffraction 
pattern as $k \Delta \theta_{\rm c.m.} \sim \pi \Delta R_{\rm m}/R_{\rm m}^2$.
In order to estimate $L$ within $\pm 20$ MeV in the present theoretical
framework, however, systematic measurements of the peak angles in the small 
momentum transfer regime for various elements are desired for neutron excesses
$\alpha\lesssim0.3$ with accuracy of $\pm$0.01 deg.

     In such an estimate of $L$, in which we focus on the angle of the first 
scattering peak, it would be useful to obtain a general relation between the 
peak angle and the parameters characterizing the adopted nucleon distributions
within the Glauber model, as obtained by Amado et al.\ \cite{ADL} 
for intermediate momentum transfers and for a Fermi distribution.  It is also 
important to examine the predictability of the peak angle by the present 
scattering model.  Neglect of the spin dependent part of the $pN$ scattering 
amplitude and of nucleon correlations, as well as uncertainties in the values 
of $\alpha_{pN}$, may affect the prediction of the peak angle and hence the
estimate of $L$, although they affect the prediction of the heights of 
diffraction minima and maxima more remarkably \cite{Alkha}.

     We finally consider proton-nucleus total reaction cross sections, which 
can be calculated from the same Glauber model as used for the elastic 
scattering calculations.  The total reaction cross section can be written as
\begin{equation}
\sigma_{\rm reac}=\int d{\bf b}\left(1-\left|e^{i\chi_{\rm N}({\bf b})}
                               \right|^2\right),
\end{equation}
where $\chi_{\rm N}$ is given by Eq.\ (\ref{phase}).  For neutron-rich stable 
nuclei such as $^{208}$Pb and $T_p\gtrsim500$ MeV, we find that the total 
reaction cross section tends to increase with either increasing $L$ or 
decreasing $K_0$.  This tendency stems from the fact that the reaction cross 
section becomes larger for larger diffuseness and radius of the nucleus.  
However, we have difficulty in deriving the EOS parameters from comparison
with the empirical reaction cross sections even if data for unstable nuclei 
are accumulated.  This is partly because errors in the empirical data 
\cite{Ray1} are of order or even larger than the change in the calculated 
reaction cross sections induced by the uncertainties in $L$ and $K_0$ and 
partly because the surface diffuseness and the tails of the nucleon 
distributions, which are underestimated and ignored in the present macroscopic
nuclear model, affect the prediction of the reaction cross sections.

     In conclusion, by expressing the nucleon distributions in a way dependent
on the EOS parameters and incorporating such distributions in the Glauber 
model, we calculated differential cross sections for proton-nucleus elastic 
scattering.  At large neutron excess, we found out a large correlation between
the peak angle in the small momentum transfer regime and the density symmetry 
coefficient $L$.  This suggests a possible method to determine $L$ from future 
systematic measurements of proton elastic scattering off unstable nuclei.

\section*{Acknowledgements}

    We are grateful to Dr.\ I. Tanihata, Dr.\ A. Kohama and Dr.\ H. Koura for 
helpful discussions.  This work was supported in part by RIKEN Special 
Postdoctoral Researchers Grant No.\ 011-52040.

%\newpage

\end{document}